\documentstyle[twoside,fleqn,espcrc2]{article}

\input eqalign.sty     

\newcommand{\Tr}{{\rm{Tr}}}

\newcommand{\be}{\begin{equation}}
\newcommand{\ee}{\end{equation}}
\newcommand{\bea}{\begin{eqnarray}}
\newcommand{\eea}{\end{eqnarray}}

\newcommand{\dds}{\stackrel{\leftrightarrow}{D}}

\begin{document}

\title{
Renormalization constants of local operators within the 
Schr\"odinger functional scheme}
\author{
A. Shindler\address{Dipartimento di Fisica, 
Universit\`a di Roma {\em Tor Vergata}}
\thanks{Presented at 
Lattice '99, Pisa, Italy.} }
\begin{abstract}
We define, within the Schr\"odinger functional (SF) scheme , the
matrix elements of the twist-2 operators corresponding to the
first two moments of non-singlet parton density, and the first moment of 
singlet parton densities. 
We perform a lattice one-loop calculation that fixes the relation 
between the SF scheme and
other common schemes and shows the main source of lattice artefacts.
Few remarks on the improvement case are added.
\end{abstract}
\maketitle
\section{Introduction}

The accurate knowledge of hadron parton densities is an essential ingredients
for the experimental text of QCD at the accelerator energies.
Their normalization is usually obtained from a fit to a set of reference
experiments and used for predicting the behaviour of hard hadron processes
in different energy regimes.
The calculation of the normalization needs non-perturbative methods.
These computation, expecially for the higher moments, \cite{roberto_proc} 
can reduce, for example, the uncertaintes on the gluon parton densities 
at values of the Bjorken $ x $ larger than $0.5$.
These calculation are made mainly by two groups (see for example 
\cite{roberto_non_pert}, \cite{schierholz_non_pert}).

It's well known that, to match the scheme of the 
non-perturbative simulation and the scheme where the 
experiment made the comparison with theory, it 
is necessary a lattice perturbative computation of the 
renormalization constants of the operator in the same scheme 
where the operator is numerically computed.

The perturbative \cite{our_pert} and non-perturbative \cite{roberto_non_pert} 
calculation are made in the Schr\"odinger functional scheme (SF).
The Schr\"odinger functional has been discussed extensively in the literature 
(see \cite{luscher_review} for reviews).
Among the advantages of
the method, we only quote the possibility of performing the
computations at zero physical
quark mass and of using non-local gauge invariant sources for fermions
and gluons without need of a gauge-fixing procedure. 
In our particular case, we exploit both features.

\section{Non-singlet}

We define the SF correlation function of the first and second 
moment of the non-singlet parton densities by the observables:
\begin{equation}\meqalign{
f_{0_{12}}(x_0;{\bf{p}}) = f_2(x_0;{\bf{p}}) = -a^6\sum_{\bf{y},\bf{z}} 
\rm{e}^{i\bf{p}(\bf{y}-\bf{z})} \times \cr
\langle \frac{1}{4} \bar\psi(x) \gamma_{[1} 
\dds_{2]}\frac{1}{2} \tau^3 \psi(x) 
\bar\zeta({\bf{y}}) \Gamma \frac{1}{2} \tau^3 \zeta({\bf{z}})\rangle \cr
f_{0_{123}}(x_0;{\bf{p}}) = f_3(x_0;{\bf{p}}) = -a^6\sum_{\bf{y},\bf{z}} 
\rm{e}^{i\bf{p}(\bf{y}-\bf{z})} \times \cr
\langle \frac{1}{8}\bar\psi(x) \gamma_{[1} \dds_{2}
\dds_{3]}\frac{1}{2}\tau^3 \psi(x)
\bar\zeta({\bf{y}}) \frac{1}{2} \Gamma \tau^3 \zeta({\bf{z}})\rangle 
\label{eq:S_observable}
}\end{equation}
\noindent where the contraction of the classical fields is 
non-vanishing if the matrix 
$\Gamma $ satisfies: $\Gamma P_{\mp} = P_{\pm} \Gamma$, where 
$P_{\pm} = \frac{1}{2} (1 \pm \gamma_0)$ and $p$ is the 
momentum of the classical field sitting on the boundary. 

The quantities $\zeta$ are the response to a variation of the classical 
Fermi field configurations on the boundaries.
We take the limit of massless quarks, but some care should be taken so as to ensure this limit at order $g^2$.
The breaking of chiral simmetry of the Wilson action entails a non-zero shift
of the quark mass from the naive value at order $g^2$.

The matrix element of the operator for the first moment involves 
two directions and three for the second moment. 
These directions must be provided by external vectors:
we have chosen to obtain one of them from the contraction matrix $\Gamma $,
i.e. from the polarization of the vector classical state $\Gamma = \gamma_2$,
and the remaining ones from the momentum $p$ of the classical Fermi 
field at the boundary.

To compute the renormalization constants of the operators is necessary to 
remove the renormalization constant of the classical boundary sources 
$\zeta$. Following ref. \cite{f_1}, this is represented by the quantity 
called $f_1$. Both $f_2$ and $f_1$ are normalized by their tree-level
expressions.
We define the renormalization constants such that operator matrix elements 
(here briefly indicated with $O$) is equal to its tree-level value at 
$\mu = 1/L$:
\begin{equation}
O^R(\mu) = Z(a \mu)^{-1} O^{\rm{bare}}(a/L)
\end{equation}
\noindent with $Z(a/L)$ defined by
\begin{equation}
O^{\rm{bare}}(a/L) = Z(a/L) O^{\rm{tree}}
\end{equation}
\noindent At the one loop the observable can be 
parametrized as:
\begin{equation}
Z(pL,x_0/L,a/L) = 1 + g^2 Z^{(1)}(a/L)
\end{equation}
\noindent with 
\begin{equation}\meqalign{
&& Z^{(1)}(a/L) = \cr 
&& b_0 + c_0 \ln(a/L) + 
\sum_{k=1}^{\infty} a^k \frac{b_k + c_k \ln(a/L)}{L^k},
\label{eq:res_parametrisation}
}\end{equation}
The study of the scale dependence of the renormalization constant
is then equivalent to that of the dependence upon the lattice size $L$, 
provided the external variables, upon which the matrix elements depend, 
scale like the basic lenght $L$.
We made the calculation with the minimum value of the momentum 
$pL = 2\pi$, and with a ``finite size'' momentum $\theta = 0.1/L$, which escape
the momentum quantization rule. Both the computation are made with 
$x_0 = L/4$, and $x_0 = L/2$.
.
The fit (the fitting procedure is described in ref. \cite{our_pert})
to the $N = L/a$ dependence of our results has been 
made by the expression
\begin{equation}\meqalign{
&& Z^{(1)}(N) = \cr 
&&B_0 + C_0 \ln(N) + \sum_{k = 1, 2} \frac{B_k + C_k \ln(N)}{N^{k}} 
\label{eq:fittype}
}\end{equation}
\noindent excluding higher-order terms.
Table \ref{tab:summ} contains a summary of the constants $B_0$ for
the operators, after removing the external legs renormalization, in
the various cases that we have discussed.
The lattice result converge much faster in their continuum limit, 
using a ``finite size'' momentum, confirming the momentum 
quantization as a major source of lattice artifacts (see ref.\cite{our_pert}).
\begin{table}
\caption{
Values of the constants of the operators renormalized
in the two definitions with the real momentum $p$ or 
the ``finite size momentum'' $\theta$ different from zero.}
\begin{tabular}{ c c c }
\hline
Moment & Definition & Constant \\
\hline
First  & $ x_0 = L/4 (p \neq 0) $ & $B_0 = 0.2635(10)$  \\
First  & $ x_0 = L/2 (p \neq 0) $ & $B_0 = 0.2762(5)$  \\
Second  & $ x_0 = L/4 (p \neq 0) $ & $B_0 = 0.1875(20)$  \\
Second  & $ x_0 = L/2 (p \neq 0) $ & $B_0 = 0.1895(50)$  \\
\hline
First  & $ x_0 = L/4 (\theta \neq 0) $ & $B_0 = 0.12180(15)$  \\
First  & $ x_0 = L/2 (\theta \neq 0) $ & $B_0 = -0.23340(15)$  \\
Second  & $ x_0 = L/4 (\theta \neq 0) $ & $B_0 = -0.06080(55)$  \\
Second  & $ x_0 = L/2 (\theta \neq 0) $ & $B_0 = -0.5675(20)$  \\
\hline
\end{tabular}
\label{tab:summ}
\end{table}
\noindent We made the same computation with clover action
\cite{our_impr}.
Also in this case lattice artefacts start 
at order $a$ for both coefficients, because 
the ${\rm O}(a)$-improvement of the action should be
complemented with the improvement of the operators and of the boundary
counterterms in order to lead to a full cancellation of effects 
appearing linear in $a$.
The Feynman rules for this computation can be easily derived using 
ref.~\cite{luscher_pert}.
The calculation is done to the order $c_{sw}^2$.
It's important to stress out that to extract the finite constant to this order
we must do the computation to the same order of $f_1$ and of 
the mass shift.
The fitting procedure is the same that in the Wilson case.
The finite part of the renormalization constant now will be defined as
\begin{equation}
B_O = B_O^{(0)} + B_O^{(1)} c_{sw} + B_O^{(2)} c_{sw}^2
\label{eq:constant}
\end{equation}
\noindent Doing the subtractions of $f_1$ and of the mass shift we obtain 
the following results:

\begin{equation}\meqalign{
&& B_O^{(1)} = -0.0327(6) \qquad  ~~~ \tilde B_O^{(1)} = -0.0327012 \cr
&& B_O^{(2)} = -0.005725(9) \qquad  \tilde B_O^{(2)} = -0.005726859
}\end{equation}
\noindent Where $\tilde B_O^{(1)}$, and $\tilde B_O^{(2)}$ 
are the coefficients computed in a standard lattice \cite{capitani_rossi}.
It's clear that the parts of the finite constant proportional to
$c_{sw}$ and to $c_{sw}^2$ are the same, within the errors, in the two schemes.
So the difference between the total finite constants in the two different 
schemes will be indipendent on the fermion action used.

\section{Singlet}

We can reproduce the same calculation in the flavour singlet sector.
In the singlet case the mixing under renormalization of the operator
force to compute $4$ correlation functions, 
involving gluonic and fermionic boundary state. 
The $2$ SF correlation function involving fermionic boundary state are:
$f_{qq}(x_0;{\bf{p}})$ which is essentialy the same of $f_2$, and
\begin{equation}\meqalign{
&& f_{gq}(x_0;{\bf p}) = \cr 
&& -a^6\sum_{\bf{y},\bf{z}} \rm{e}^
{i\bf{p}(\bf{y}-\bf{z})}
\langle \Tr \{ \rm{F}_{1 \rho} \rm{F}_{ \rho 2} \} 
\bar\zeta({\bf{y}})\Gamma \zeta({\bf{z}})  \rangle
}\end{equation}
To study the other two correlation functions it is necessary to fix
the gluonic boundary field.
We use again the particular gauge group that leaves invariant the 
Schr\"odinger functional.
We define the $\em{Big~Teeth}$ state as follows
\begin{equation}\meqalign{
&& D_i( {\bf y}) = \{ \prod_{k=0}^{\frac{N}{4} -1} U_0 ({\bf y} +ka \hat 0) \}
U_i ({\bf y} + \frac{T}{4} \hat 0) \times \cr 
&& 
\{ \prod^{0}_{k=\frac{N}{4} -1} U_0^{-1} ({\bf y} + a \hat i + k a \hat 0) \}  
}\end{equation}
\noindent We define then the gluonic boundary state as
\begin{equation}
G_{12}({\bf y},{\bf z}) = \Tr [g_1({\bf y}) g_2({\bf z})]
\end{equation}
\noindent where
\begin{equation}
g_i({\bf y}) = \frac{1}{2 i a g_0} \{ D_i( {\bf y}) - D_i( {\bf y})^{-1} \}
\end{equation}
The $2$ remaining correlation function are defined as:
\begin{equation}\meqalign{
&& f_{qg}(x_0) = \cr
&& -a^6 \sum_{\bf{y},\bf{z}} 
\langle \frac{1}{4} \bar\psi(x) \gamma_{[1} 
\dds_{2]} \psi(x) 
G_{12}({\bf y},{\bf z})  \rangle
}\end{equation}
\begin{equation}
f_{gg}(x_0) = a^6 \sum_{{\bf y},{\bf z}} 
\langle \Tr \{ \rm{F}_{1 \rho} \rm{F}_{ \rho 2} \}
G_{12}({\bf y},{\bf z}) \rangle
\end{equation}
\noindent It's important to note that the SF scheme allow us to construct 
a gauge invariant gluonic source projected at $0$ momentum with a tree-level 
well defined in the continuum limit. This is achieved by the two different 
polarization of the gluonic source. This improve the convergence to 
the continuum limit as it's been remarked
in the non-singlet calculation.  
The results for $f_{gq}$ and $f_{qg}$ are the following:
\begin{equation}
B_{gq}^{(0)} = 0.0261(3) \qquad B_{qg}^{(0)} = 0.02919(1)
\end{equation}
\noindent It's currently under way the computation of $f_{gg}$ and of the 
gluonic source renormalization.



\def\NPB #1 #2 #3 {Nucl.~Phys.~{\bf#1} (#2)\ #3}
\def\NPBproc #1 #2 #3 {Nucl.~Phys.~B (Proc. Suppl.) {\bf#1} (#2)\ #3}
\def\PRD #1 #2 #3 {Phys.~Rev.~{\bf#1} (#2)\ #3}
\def\PLB #1 #2 #3 {Phys.~Lett.~{\bf#1} (#2)\ #3}
\def\PRL #1 #2 #3 {Phys.~Rev.~Lett.~{\bf#1} (#2)\ #3}
\def\PR  #1 #2 #3 {Phys.~Rep.~{\bf#1} (#2)\ #3}

\def\etal{{\it et al.}}
\def\ibid{{\it ibid}.}

\end{document}